\documentclass[twocolumn,showpacs,preprintnumbers,amsmath,amssymb, nofootinbib]{revtex4}

\usepackage{pdflscape}
\usepackage{color}
\usepackage{dcolumn}
\usepackage{bm}
\usepackage{epsfig}%
\usepackage{amsmath}
\usepackage[colorlinks]{hyperref}
\usepackage{ulem}

\begin{document}

\title{Microscopic parameters of a type-II superconductor measured by small-angle neutron scattering}

\author{D. Alba Venero$^1$, A.-M. Valente-Feliciano$^2$, O. O. Bernal$^3$ and V. Kozhevnikov$^4$}
\affiliation{
$^1$ISIS Neutron and Muon Source, Appleton and Rutherford Laboratory, OX11 0QX, UK \\
$^2$Thomas Jefferson National Accelerator Facility, Newport News, VA 23606, USA  \\
$^3$California State University LA, CA 90032, USA\\
$^4$Tulsa Community College, Tulsa, OK 74119, USA}.  \\


\begin{abstract}
\noindent A necessary condition for understanding and predicting the properties of any material is knowledge of microscopic parameters which control these properties in a state of thermodynamic equilibrium. 
One can show (see, e.g., Ref.\,\cite{VK_book}), that in superconductors these parameters are the radius of the orbital motion of electrons bound in Cooper pairs $R_0$ and the radius of the field-induced currents $r_i$ caused by precession of the pairs; one more parameter, associated with $r_i$, is the number density of Cooper pairs $n_{cp}$. In this paper  we report on the first measurements of these parameters in a type-II superconductor (niobium) by SANS (small-angle neutron scattering). Other approaches  to measuring  the microscopic parameters are considered. Our work suggests  novel avenues for studying superconductivity, 
important for disclosing its mechanisms in superconductors of all kinds. 
\end{abstract}\

\maketitle

\textbf{Theoretical introduction.}
\vspace{1mm}

 In the superconducting (S) state, a statistically significant proportion of conduction electrons are combined 
 into Cooper pairs (CPs), which are stable formations of two 
 electrons with (1) mutually compensated spins and (2) zero net linear momentums \cite{Cooper}. 

While the first condition ensures 
the pairs stability and thermodynamic advantage of the S state (since the free energy of spinless units is always  less than the free energy of the same units with spins \cite{Landau_QM}), the second implies that the paired electrons orbit their centers of mass. The latter means that each CP possesses an angular momentum $\bm{\iota}_0$ with a corresponding magnetic moment  $\bm{\mu}_0=\gamma\bm{\iota}_0$, where $\gamma(= ge/2mc$ with $e$, $m$ and $c$ standing for the unit charge, a free electron mass and the speed of light, respectively) is the gyromagnetic ratio of the orbiting electrons. Accordingly, their $g$-factor is unity, as was first measured by I. Kikoin and Goobar  \cite{Isaak-1,Isaak-2}. 

\begin{figure}
	\centering
	\includegraphics[width=0.95\linewidth]{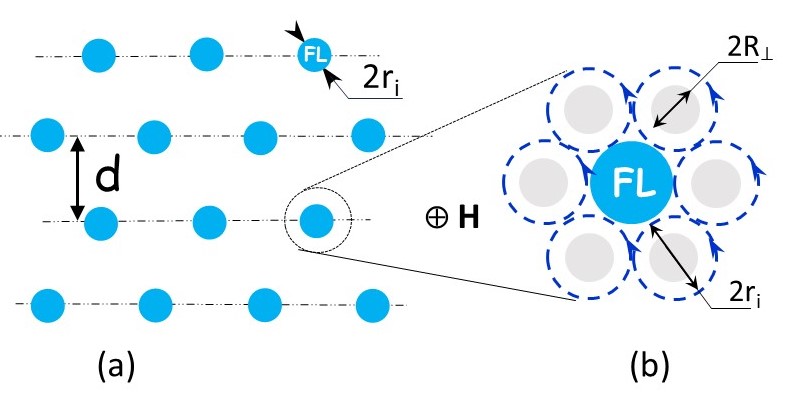} 
	\caption{ (a) Equilibrium structure of the flux lines (FLs) in type-II superconductors in the mixed state. (b) Zoomed area with a single FL passing through the network of micro-whirls; the FL takes the space originally occupied by the micro-whirl. $R_\perp$  is the rms radius of cylindrical volumes filled with CPs of different orientations; $r_i$  is the rms radius of micro-whirls equal to the radius of the field-induced currents in CPs and, respectively, to the radius of the FLs; $d$ is a parameter of the FL lattice (distance between neighboring scattering planes). Arrows in (b) depict induced currents.  
	$\textbf{H}$ is the intensity of the magnetic field directed into the page.} 
	\label{fig:epsart}
\end{figure}

Therefore, being in the field $\textbf{H}$ (unlike the induction $\textbf{B}$ of a magnetic field, its intensity or strength $\textbf{H}$ is never zero \cite{Maxwell})  
CPs precess with Larmor frequency $\bm{o}=-\gamma\textbf{H}$, thus forming field-induced circular current loops residing in transverse to $\textbf{H}$ planes. Note that Larmor precession is the only kind of electron motion in magnetic systems, which meets the Second law of thermodynamics for equilibrium states, namely the invariance of the systems' properties with respect to reversal of time \cite{Planck}.

A squared rms radius of the loops is $r_i^2=m_{cp}c^2/\pi n_{cp}(2e)^2$, where $m_{cp}$ and $n_{cp}$ are mass and number density of CPs \cite{ri vs lambda} (see Appendix).  Due to its origin (Larmor precession), $r_i$ does not depend on the field, as it also follows from the experiment  \cite{Pippard-1950,VK_JSNM,muSR-2025}. Each loop, i.e., the current induced in each CP, has a negative magnetic moment. This leads to ideal diamagnetism of the S-phase in thermodynamically equilibrium samples, as was discovered independently by Meissner and Ochsenfeld \cite{Meissner and Ochsenfeld}, and  by Ryabinin and Shubnikov \cite{Shubnikov-Rjabinin}.

\begin{figure}
	\includegraphics[width=0.9\linewidth]{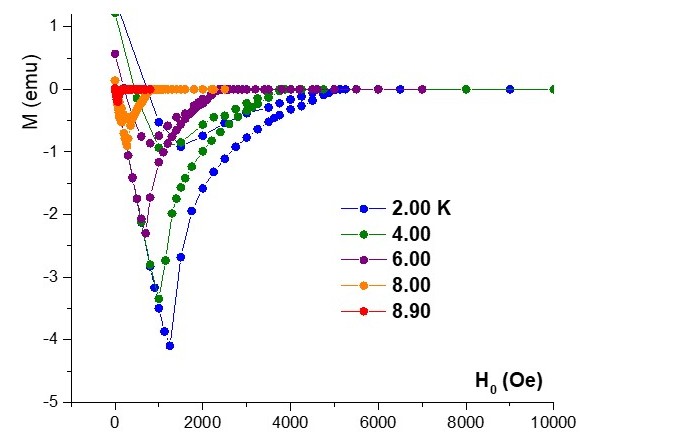} 
	\caption{Dependence of the magnetic moment $\textbf{M}$ on the applied field $\textbf{H}_0$ at constant temperatures, as indicated.  The sample/field configuration was identical to that shown in Fig.\,3. The measurements were performed on a 7 mm diameter sample, which was a twin of the sample used in the SANS experiment (fabricated from the same Nb rod and subjected to the same treatment). The sample was cooled in zero field. At an ascending field (in the zero-field-cooled (ZFC) mode), the sample is in thermodynamic equilibrium, while at a descending field, or in the FC mode, it is away from equilibrium.}
	\label{fig:epsart}
\end{figure}

To minimize the free energy of the ensemble of CPs, the induced currents line up in an ideally ordered 2D hexagonal lattice of micro-whirls oriented parallel to $\textbf{H}$. Hence, the entropy of the pairs’ ensemble is zero, which is confirmed  by the absence of thermoelectric effects in superconductors, first revealed by Meissner \cite{Meissner27}, as well as by the absence of temperature dependence of magnetization in diamagnets.
In turn, zero entropy implies zero temperature of the ensemble of ordered  pairs, one of the consequences of which is zero resistivity of the S phase at subcritical total current \cite{zero Tcp}. 

The micro-whirls represent long solenoids of radius $r_i$ very tightly "wound`` around the H-lines (a longitudinal distance  between the neighboring current loops equals  $2e^2/mc^2\approx 5\cdot 10^{-13}$ cm, about tripled diameter of the proton). Consequently, the interaction between neighboring loops in a micro-whirl is mutually compensated and the micro-whirls do not interact with each other, in accordance with the requirements of thermodynamics. 

On the other hand, in the case of inhomogeneous magnetization, the FLs pass through the sample via normally conducting holes in the network of superconducting micro-whirls.  Herewith geometry of the FLs in type-II superconductors reproduces the geometry of the original micro-whirls' structure, which also follows from thermodynamics. The cross-section of the equilibrium structure of the FLs and a zoomed area with one FL are shown in Fig\,1. To appreciate the degree of FL ordering, note that number density of FLs is of the order of $10^8$ mm$^{-2}$.

The radius of the orbital motion of paired electrons $R_0$, more specifically, its rms projection onto the transverse to $\textbf{H}$ plane $R_\perp=R_0\sqrt{2/3}$ (see Appendix), and the rms radius of the field-induced currents $r_i$ are key microscopic parameters controlling properties of superconductors in a state of thermodynamic equilibrium. In particular, if $r_i>R_\perp$, then the superconductor is of type II and vice versa \cite{Note on IMS}. At the same time, the ratio $r_i/R_\perp\equiv\aleph$  is a constant of material depending on its purity. One can show that in type-II materials $\aleph=H_{c2}/H_c$, where $H_c$ is the thermodynamic critical field. To measure these microscopic parameters was the purpose of our study. 

The outlined micro-whirls model (MWM) 
is discussed in detail in \cite{VK_book}; its shortened version is available in \cite{Encyclopedia}. The original account of the MWM and experimental data that triggered  
its development are presented in \cite{VK_JSNM}. The MWM  is based on the Bohr–Sommerfeld quantization condition applied to Cooper pairs; it is free of postulates and adjustable parameters. The MWM consistently describes all known properties of superconductors and predicts not yet known effects. The idea of micro-whirls in superconductors was first put forward by Hall \cite{Hall}.  
\vspace{3mm}	 

\begin{figure}
	\includegraphics[width=0.81\linewidth]{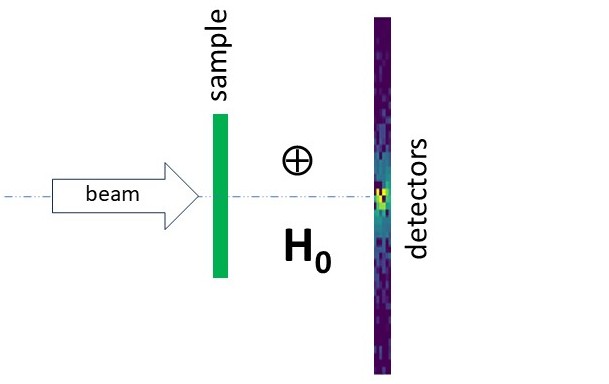} 
	\caption{The sample/field configuration in the SANS experiment.  $\textbf{H}_0$ is the applied field directed parallel to the sample plane.   } 
	\label{fig:epsart}
\end{figure}

\begin{figure*}
	\includegraphics[width=1.0\linewidth]{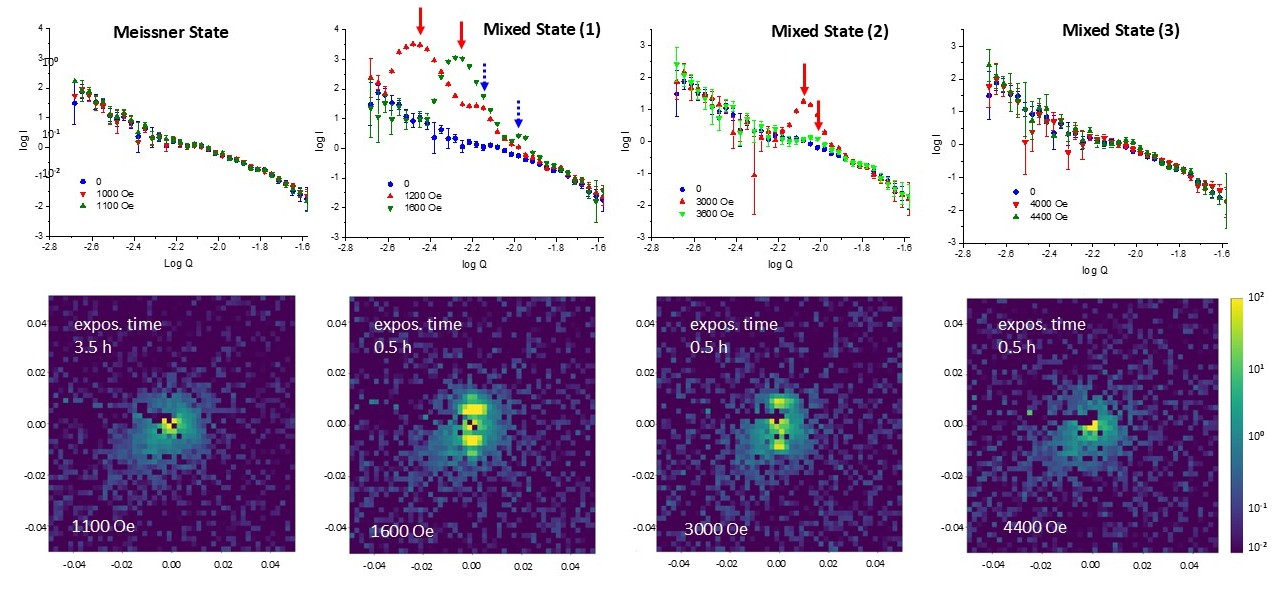} 
	\caption{Experimental data and Laue-grams for the scattered neutron intensity $I$ (arb. units) vs the scattering vector $Q$ (1/$\mathring{A}$) at $T=3.5$ K and different applied fields. Solid (dashed) arrows mark the $Q$-vectors at the maxima of the first (second) order. } 
	\label{fig:epsart}
\end{figure*}

\textbf{Experimental}
\vspace{1mm}

According to an estimate based on Low-Energy $\mu$SR data \cite{muSR-2025,VK_JSNM}, 
the diameter of micro-whirls in Nb at $T\approx$ 2 K is about 100 nm. Hence, in principle, $r_i$ can be measured by SANS (small-angle neutron scattering). This is the experiment we performed using a polychromatic beam of unpolarized neutrons on ZOOM instrument of the ISIS Neutron and Muon Source. The chief challenge was the low contrast of the induced currents against the background of arbitrarily oriented orbital currents, since the former are orders of magnitude smaller than the latter, as is  always the case with Larmor precession.

The sample was a one-side polished 1 mm thick single crystal Nb disc 19 mm in diameter fabricated in the Surface preparation laboratory, the Netherlands. It was annealed at 800 \textcelsius\,   for 3 hours and electropolished. The applied magnetic field was parallel to the stationary sample and perpendicular to the neutron beam.  The sample was cooled in a zero field, then the field was turned on and varied in one direction. At such conditions the flux trapped in the sample is negligible and therefore, as was first shown by Shubnikov with coworkers \cite{Shubnikov-37_typeII,Shubnikov-Rjabinin}, the sample of our geometry is in thermodynamic equilibrium even if its purity is not very high. 

The magnetization curves measured on a 7-mm in diameter twin's sample and the sample/field configuration are shown in Figs.\,2 and 3, respectively. As can be seen from Fig.\,2, the magnetization data obtained in the ZFC mode are reproducible; in particular, it is easy to recognize that in the linear section of the magnetization curves the sample is in the Meissner state. This confirms the thermodynamic equilibrium of our sample in ZFC mode. Contrarily, in FC mode, i.e, at the descending field, the sample is out of equilibrium. Note that most previous SANS experiments were performed with FC samples of  transverse geometry \cite{Sebastian}, where thermodynamic equilibrium is practically unattainable. 

The measurements were performed at 3.5 K; the critical fields $H_{c1}$, $H_c$ and $H_{c2}$ in our sample 
at this temperature are  1100, 1800 and  4100 Oe, respectively, with an error within 100 Oe for  $H_{c1}$ and 50 Oe for $H_c$ and $H_{c2}$. All critical fields were determined from magnetization data shown in Fig.\,2. The thermodynamic critical field $H_c$ was calculated from definition: $H_c^2 V=-8\pi\int \textbf{M}\cdot d\textbf{H}_0$, where $V$ is the sample volume and the integral is taken over the entire field range of the S state at constant temperatures.
\vspace{2mm}

\textbf{Results and discussion.}
\vspace{1mm}

SANS patterns obtained on the sample in the Meissner state and typical patters of the mixed state along with corresponding Laue-grams are shown in Fig.\,4. 

The data taken in the Meissner state are presented in the left panels of Fig.\,4. The exposition time was 3.5 hours per data-point. No ordered structure was detected. Consequently, at this point  we are unable to estimate the minimum 
neutron flux needed to detect the current structure of the Meissner state, nor whether  neutrons are applicable for this task in general. 

In the mixed state the situation was more favorable. As seen from  Fig.\,4, 
we observed three kinds of the diffraction patterns, with maxima of the first and second order, one maximum and no maximum.  The exposition time per data-point was half an hour. Experimental points for the intensity vs the scattering vector (log\,$I$ vs log\,$Q$) in vicinity of each maximum were approximated with the Gaussian curve from which the values of scattering vectors at the maxima  $Q_{m1}$ and $Q_{m2}$, and one-sigma error bars were evaluated. Positions of $Q_{m1}$ and $Q_{m2}$ vs the applied field $H_0$, and calculated from them the FL lattice
 parameter $d$  are shown in Fig.\,5.

After passing $H_{c1}$ from below the FLs start to fill the sample bulk, implying that 
 more and more micro-whirls are replaced by FLs, and the lattice parameter $d$ decreases accordingly.   
As follows from Fig.\,5b, the process is not linear: at first it resembles an avalanche, then continues with a gradually decreasing pace. This is consistent with the shape of the magnetization curves in Fig.\,2 and with 
results of other authors, including those first reported by Shubnikov at al. \cite{Shubnikov-37_typeII}; the field dependence of the diffraction pattern (Fig.\,5a) and the dynamics of observed variation of the lattice parameter (Fig.\,5b) are consistent with SANS and STM data reported by Christen et al. \cite{Christen} 
and Hess et al. \cite{Hess}, respectively. 

At some point the FLs begin merging, which leads to gradual destruction of the FL lattice and eventually to disappearance of the diffraction maximum.  Taking into account the hexagonal geometry of the lattice, this means that when $H_0 \rightarrow H_{c2}$, $d$ tends to its minimum,  equal to $r_i\sqrt{3}$. According to the data shown in Fig.\,5b, the minimum $d$ is equal to $710 \pm 100\,\mathring{A}$ or $r_i=41\pm6$ nm. 
This value of $r_i$ is consisted with the aforementioned estimate following from the Low-Energy $\mu$SR data.    

\begin{figure}
	\includegraphics[width=0.95\linewidth]{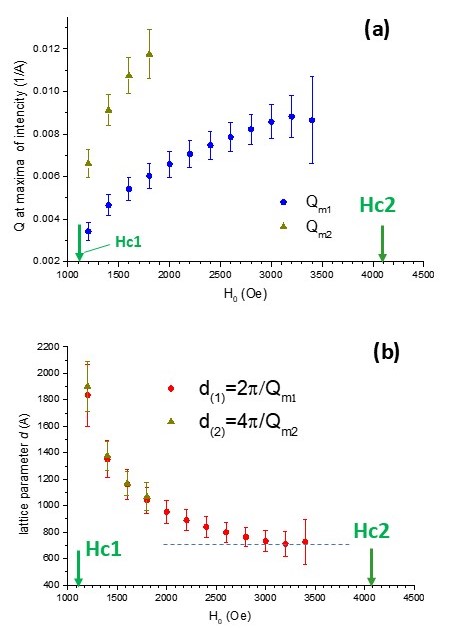} 
	\caption{(a) Scattering vector at maxima of the first ($Q_{m1}$) and second ($Q_{m2}$) orders. (b) The flux line lattice parameter $d$ calculated from $Q_{m1}$ and $Q_{m2}$. The dashed line designates the minimum $d$. $H_0$ is the applied field.  } 
	\label{fig:epsart}
\end{figure}

Next, having $r_i$ one can compute the number density of Cooper pairs $n_{cp}=m_{cp}c^2/\pi r_i^2 (2e)^2=mc^2/2\pi r_i^2e^2$, where $m$ is the free electron mass. This yields $n_{cp}=(3.3\pm0.8)\cdot 10^{22}$ cm$^{-3}$. Neglecting thermal expansion of Nb, number density of free (non-paired) electrons in this metal at zero temperature is  $n_e=5.5\cdot 10^{22}$ cm$^{-3}$ \cite{Grigoriev}. Hence, $n_{cp}/n_e=0.6\pm0.15=60\pm15$\,\%. Thus, our data furnish a confirmation of the hypothesis of Gorter and Casimir of 1934 that at zero temperature all conduction electrons are in a condensed, i.e. paired, state \cite{Gorter_Casimir}. Note that if, as someone might assume, 
$r_i$ is equal to the London penetration depth, then $n_{cp}$ would be four times smaller, which is quite unnatural.

Finally, the radius of the orbital motion of the paired electrons $R_0$ can be computed from the above definition of $\aleph$ and its relationship with $H_c$ and $H_{c2}$ for type-II superconductors. From this follows that  $R_0=\sqrt{3/2}\,r_i/\aleph=\sqrt{3/2}\,r_i {H_c}/{H_{c2}}$. Hence, $R_0$ in Nb at 3.5 K equals $22\pm 5$ nm. It is worth noting that $R_0$ is the most ``hidden" microscopic parameter and, accordingly, the least accessible  for direct measurements (if such measurements are possible at all). On the other hand, according to the MWM, $R_0$ is the most fundamental parameter, since the other two ($r_i$ and $n_{cp}$) depend on the sample purity. 
\vspace{2mm}

\textbf{Summary and outlook}.
\vspace{1mm}
 
We demonstrated an approach allowing to determine equilibrium microscopic parameters of type-II superconductors from SANS patterns. The main advantage of this approach is that data obtained are conditioned by the properties of the sample bulk. Accordingly, the data retrieved from SANS are indifferent to the state of the sample surface and therefore allow an objective assessment of the error.

The parameters reported are the radius of the field-induced microscopic currents in Cooper pairs, related to it the number density of the pairs, and the most ``sacred" parameter, the radius of the orbital motion of the paired elections, i.e., the radius of Cooper pairs. All these values were acquired for the first time. The accuracy of the obtained quantities  
is within 25\%; it can be improved by enlarging the exposure time. 

We were unable to evaluate a possibility of measuring the microscopic parameters on a sample in the Meissner state. SANS experiments targeted to this goal should be carried out with substantially longer exposure time (probably on the order of a day) close to the field $H_{c1}$ at the lowest achievable sample temperature; the use of polarized neutrons should be a plus. 

Alternatively, the microscopic parameters can be determined in a lesser time-consuming way from ESR and NMR measurements with homogeneous samples, i.e., with the samples in the Meissner state. To date, none of these resonances have been studied at such conditions. As expected, ESR spectra may have two resonances associated with precession of the bound and free electrons with $g$-factors 1 and 2, respectively. A primary parameter which can be extracted from these spectra is the number density of Cooper pairs $n_{cp}$. From there one can compute $r_i$ and, for type-II materials, $R_0$. The sample size in such cases may be significantly smaller than that required for SANS (but in any case the shape should be as ellipsoidal as possible).  

To find $R_0$ in type-I superconductors, one can use the approach applied by Shubnikov et al. \cite{Shubnikov-37_typeII} and Pippard \cite{Pippard-53}.  It consists in transforming a type-I superconductor into a type-II one by admixing an original type-I metal with a small amount ($\sim$ 0.1-1 wt.\%) of another metal so that the condensation energy $E_c(=-\int \textbf{M}\cdot d\textbf{H})$ remains unchanged. 
\vspace{2mm}

\hspace{-3mm}\textbf{Acknowledgment.} We are grateful to Dr. Sebastian  M\"{u}hlbauer for helpful discussions.

\vspace{2mm}
\hspace{-3mm}\textit{Data availability} 

\hspace{-3mm}The original data are available in the ISIS repository at 
https://doi.org/10.5286/ISIS.E.RB2220299-4. 

\vspace{2mm}
\textbf{Appendix.} \textit{Microscopic parameters $r_i$ and $R_\perp$.} 
\vspace{2mm}

As indicated in the main text, a detailed discussion of formulas mentioned in this paper is available in \cite{VK_book} and \cite{VK_JSNM,Encyclopedia}. Below we provide a concise derivation of the formulas for the microscopic parameters $r_i$ (the radius of the field induced current in CPs) and $R_\perp$ (the rms projection of the CP's radius $R_0$ onto a transverse to  $\textbf{H}$ plane). 

Before we begin, we need to refresh the notions of the induction \textbf{B} and the field intensity, or strength, \textbf{H} related as $\textbf{B}=\textbf{H}+4\pi \textbf{I}$, where $\textbf{I}$ is magnetization. 
After Lorentz \cite{Lorentz}, $\textbf{B}$ is a mean field due to all microscopic currents in the sample. In turn, $\textbf{I}$ is related to the field via the \textit{fundamental equation of induced magnetism} \cite{Maxwell}, which is
\begin{equation*}
\textbf{I}=\chi\textbf{H},\qquad(A1)
\end{equation*}
where $\chi$ is the magnetic susceptibility. In the S phase of superconductors $\chi=-1/4\pi$. 

Therefore, the intensity $\textbf{H}$, also called the magnetizing force \cite{Maxwell}, controls magnetization, i.e., it generates a magnetic moment in the microscopic units composing the sample. 
In other words,  $\textbf{H}$ is a mean field acting on a given microscopic unit; it does not include the field due to the unit in question, while $\textbf{B}$, the mean field due to all units, includes it. This seemingly minor difference between $\textbf{H}$ and $\textbf{B}$ fields inside the sample (outside $\textbf{B}\equiv \textbf{H}$) leads to principally different boundary conditions for these fields \cite{LL_ECM}. Accordingly, $\textbf{H}$ and $\textbf{B}$ \textit{must be carefully distinguished} \cite{Maxwell}. 
Besides, in superconductors the difference is by no means minor: in the S phase $\textbf{B}=0$, while $\textbf{H}$ is equal to or greater than the applied field $\textbf{H}_0$, e.g., in cylindrical superconductors in the Meissner state placed in the transversely applied field, the intensity $\textbf{H}=2\,\textbf{H}_0$. 

Like any other magnetic field, the $\textbf{H}$ field is solenoidal  (see \cite{VK_book} for details).  Therefore, its physically adequate description is based on a vector potential $\textbf{A}$ defined so as $\textbf{H}=\nabla \times \textbf{A}$ 
with Coulomb gauge ($\nabla \cdot \textbf{A}=0$) presumed automatically \cite{Tamm}.  Herewith, $\textbf{A}$ should be taken in the circular gauge (i.e., $\textbf{A}=\textbf{H}\times \textbf{r}/2$, where $\textbf{r}$ is the radius-vector lying in the transverse to \textbf{H} plane) since such a gauge is the only one complying with the cylindrical symmetry of the uniform field \textbf{H} (all directions in the planes transverse to $\textbf{H}$ are equivalent). 

Consider, for simplicity, a cylindrical superconducting sample in a parallel field $\textbf{H}_0<\textbf{H}_{c1}$. Such a sample is in the Meissner state and the intensity $\textbf{H}$ inside it equals $\textbf{H}_0$ due to the boundary condition. Magnetic moment of this sample is
\begin{equation*}
\textbf{M}\equiv V\textbf{I}=-(V/4\pi)\textbf{H}=-(V/4\pi)\textbf{H}_0 \qquad(A2)
\end{equation*}
and its magnetic energy $E_m$, i.e., the energy it takes to magnetize the sample \cite{LL_ECM}, is  
\begin{equation*}
	E_m\equiv -\int \textbf{M}\cdot d\textbf{H}_0=(V/8\pi)H_0^2=(V/8\pi)H^2. \qquad(A3)
\end{equation*}

One more important formula that we will need later is related to the sample magnetic moment at zero field:  
\begin{equation*}
\textbf{M}_0=\Sigma \bm{\mu}_0=0, \qquad(A4)
\end{equation*}
where $\bm{\mu}_0$ is magnetic moment of an individual CP caused by the orbital motion of paired electrons about their center of mass and summation is taken over all CPs of the sample. This formula means that the angular distribution of the moments $\bm{\mu}_0$ possesses a spherical symmetry. 

When $\textbf{H}_0$ is turned on, the intensity inside the sample changes from zero to $\textbf{H}$ and each electron in CPs experiences an action of the Lorentz force $\textbf{F}_L$ in addition to the forces acting at zero field. (The action of the Lorentz force on the nuclei and all other electrons produces a feeble dia- or paramagnetic response, which is neglected.)   In superconductors the electrostatic field is screened by the free (non-paired) electrons, so the Lorentz force is 
\begin{equation*}
		\textbf{F}_L=-\frac{e}{c}\left(\frac{\partial \textbf{A}}{\partial t}\right)+\frac{e}{c}\textbf{v}\times \textbf{H}, 
		\qquad(A5)
\end{equation*}   
where $\textbf{A}$ is the vector potential of the $\textbf{H}$-field, $\textbf{v}$ is the electron velocity and $t$ is rime. 

The second term in Eq.\,(A5) is the magnetic force $\textbf{F}_M$  which is transverse to velocity. Therefore, $\textbf{F}_M$ does not change the velocity magnitude, while  its action on the paired electrons  produces a torque $\bm{\tau}=\bm{\mu}_0\times \textbf{H}$. Consequently, due to non-zero $\bm{\iota}_0$, the magnetic moments of all CP precess about the direction of $\textbf{H}$. Precession starts simultaneously all over the sample with Larmor frequency, 
which does not depend on parameters of the electron orbits, and themselves parameters (the orbital radius $R_0$ and  the moment $|\bm\mu_0|$) are unaffected by the field, as is always the case with Larmor precession \cite{VK_book}. Hence, all CPs precess synchronously, retaining their mutual orientation they had in zero field and  Eq.\,(A4) holds regardless of the field. Therefore,  $\textbf{F}_M$ \textit{does not} induce magnetization. In other words. $\textbf{F}_M$ alone cannot create magnetism, which is consistent to the fact that $\textbf{F}_M$ does not do work.

Magnetization   of superconductors (as well as normal diamagnets) is due to the first term of the Lorentz force in Eq.\,(A5). This is the electric force $\textbf{F}_E$ caused by the induced vortex electric field existing while the magnetic field is changing.  $\textbf{F}_E$ performs work which changes the kinetic energy of CPs and produces the sample magnetic moment. The Newton's acceleration law for this force  is

\begin{equation*}
	\textbf{F}_E=m\frac{\partial \textbf{v}}{\partial t}=-\frac{e}{c}\left(\frac{\partial \textbf{A}}{\partial t}\right). \qquad(A6)
\end{equation*}  

The time in Eq.\,(A6) cancels out, after which it becomes
\begin{equation*}
	d\textbf{v}=-\frac{e}{mc}d\textbf{A}. \qquad(A7)
\end{equation*}

Upon integration, one finds the change in the electrons velocity caused by the magnetic field $\textbf{v}_i = \textbf{v}'-\textbf{v}_0$, where $\textbf{\textbf{v}}_0$ and $\textbf{v}'$ are the velocities before and after the field was turned on, respectively. So, $\textbf{v}_i$, the field-induced velocity of electrons, is 
\begin{equation*}
	\textbf{v}_i=-\frac{e}{mc}\textbf{A}=-\frac{e}{2mc}\textbf{H}\times \textbf{r}, \qquad(A8)
\end{equation*}
where $\textbf{r}$ is the radius of the induced circular motion centered in the pairs' centers of mass.

Accordingly, the change of angular velocity of CPs caused by the field, or the field-induced angular velocity $\bm{\omega}_i(=\textbf{r}\times \textbf{v}_i/r^2)$, is 
\begin{equation*}
	\bm{\omega}_i=-\frac{e}{2mc}\textbf{H}. \qquad(A9)
\end{equation*}

Thus, the induced angular velocity of the CP $\bm{\omega}_i$ is equal to the Larmor frequency $\bm{o}$. Similarly, it is easy to show that any change in the intensity $\textbf{H}$ leads to such a change in the angular velocity of CPs  that their final induced frequency $\bm{\omega}_i$ is equal to the Larmor frequency $\bm{o}$. In other words, CPs constantly precess with frequency $\bm{o}$.

As follows from Eq.\,(A9), $\bm{\omega}_i$ (and therefore the induced current $J_{cp}$) is independent on $r$. However, it is not the case for the induced magnetic moment $\bm{\mu}_{cp}$ and the change of kinetic energy of CPs $\epsilon_{cp}$:  they both depend on $r^2$. However, $r$ of differently oriented orbits (relative to $\textbf{H}$) are different. So, to calculate the induced magnetic moment of the sample we have to use the mean square of the radius $\left<r^2\right>$, denoted as $r_i^2$. Then, using Eq.\,(A9), an average induced  magnetic moment per CP is
\begin{equation*}
	\bm{\mu}_{cp}=\frac{\pi r_i^2}{c}\,\frac{(2e)}{2\pi}\bm{\omega}_i=\frac{\pi r_i^2}{c}\,\frac{(2e)}{2\pi}\bm{o}=-\frac{(2e)^2r_i^2}{4m_{cp}c^2}\textbf{H}, \qquad(A10)
\end{equation*}
where $m_{cp}=2m$.

In a uniform $\textbf{H}$, the magnetic moment of the sample is   
\begin{equation*}
	\textbf{M}=Vn_{cp}\bm{\mu}_{cp}=-Vn_{cp}\frac{(2e)^2r_i^2}{4m_{cp}c^2}\textbf{H}. \qquad(A11)
\end{equation*}

Comparing this expression with Eq.(A2), one finds 
\begin{equation*}
	r_i^2=\frac{m_{cp}c^2}{\pi n_{cp}(2e)^2}. \qquad(A12)
\end{equation*}

The same result follows from calculating the field-induced kinetic energy of the CPs and comparing it with Eq.\,(A3). Indeed, the change of kinetic energy per CP caused by the field is
\begin{equation*}
	\epsilon_{cp}=2\frac{mv_i^2}{2}=m\left(\frac{e}{2mc}Hr_i\right)^2=\frac{E_m}{Vn_{cp}}=\frac{H^2}{8\pi  n_{cp}}, \qquad(A13)
\end{equation*}
where the expression in parentheses is $\textbf{v}_i$ from Eq.\,(A8). 

Comparing the third and last expressions, one obtains Eq.\,(A12).

All this means that (a) the micro-whirls model meets the law of energy conservation and (b) the induced currents do not interact; the explanation of this  fact is available in \cite{VK_JSNM,Encyclopedia,VK_book}.

After all, the spherical symmetry of the distribution of the magnetic moments $\bm{\mu}_0$ means that $\left<x^2\right>=\left<y^2\right>=\left<z^2\right>=R_0^2/3$, where $x$, $y$ and $z$ are rms values of coordinates of the orbiting paired electrons with respect to their center of mass, and $R_0$ is the radius of CPs. Therefore, the mean square value of the projection of $R_0$ onto a plane transverse to $\textbf{H}$ is  $R_{\perp}^2=(2/3)R_0^2$.




\end{document}